\documentclass[10pt,emptycopyrightspace]{ewsn-proc}

\usepackage[nolist]{acronym}
\usepackage[binary-units=true]{siunitx}
\PassOptionsToPackage{hyphens}{url}
\usepackage{hyperref}
\hypersetup{
	colorlinks,
	linkcolor={blue!45!black},
	citecolor={black},
	urlcolor={black}
}
\usepackage{xspace}
\usepackage{tcolorbox} \usepackage[nolist]{acronym}
\usepackage{tabularx}
\usepackage{readarray}
\usepackage{booktabs}
\usepackage{listings,multicol}
\def\sectionautorefname{\S}
\def\subsectionautorefname{\S}

\newcommand{\shortautoref}[1]{\begingroup \renewcommand\sectionautorefname{\S}\renewcommand\subsectionautorefname{\S}\renewcommand\subsubsectionautorefname{\S}\autoref{#1}\endgroup
}

\newcommand{\sname}{ScaleClock\xspace}
\newcommand{\etal}{\textit{et al.}~}
\newcommand{\eg}{\textit{e.g.,}~}
\newcommand{\ie}{\textit{i.e.,}~}
\newcommand{\etc}{\textit{etc.}\xspace}
\newcommand{\one}{({\em i})\xspace}
\newcommand{\two}{({\em ii})\xspace}
\newcommand{\three}{({\em iii})\xspace}
\newcommand{\four}{({\em iv})\xspace}

\makeatletter
\renewcommand{\keywords}[1]{\paragraph{Keywords} #1}
\renewcommand{\paragraph}[1]{\vspace*{0.03in}\noindent{\bf #1.}\hspace{0.25ex \@plus1ex \@minus.2ex}}
\renewcommand{\subsubsection}[1]{\vspace*{0.03in}\noindent{\bf #1.}\hspace{0.25ex \@plus1ex \@minus.2ex}}
\newcommand{\paragraphS}[1]{\vspace*{0.03in}\noindent{\bf #1}\hspace{0.25ex \@plus1ex \@minus.2ex}}
\makeatother

\newcommand{\postfigspace}{\vspace{-0.28cm}}

\usepackage{amsmath,amsfonts}
\usepackage{cite}
\usepackage{algorithmic}
\usepackage{array}
\usepackage[font=small,labelformat=simple]{subfig}

\usepackage{textcomp}
\usepackage{stfloats}
\usepackage{url}
\usepackage{verbatim}
\usepackage{graphicx}
\def\BibTeX{{\rm B\kern-.05em{\sc i\kern-.025em b}\kern-.08em
		T\kern-.1667em\lower.7ex\hbox{E}\kern-.125emX}}
\usepackage{balance}
\usepackage{listings,mdframed}

\newif\ifsqueeze
\squeezefalse

\ifsqueeze
\usepackage{etoolbox}
\makeatletter
\patchcmd{\@maketitle}
{\advance\dimen0 by -12.75pc\relax}
{}
{}
{}
\patchcmd{\@maketitle}
{\end{center}}
{\end{center}\vspace{-0.05em}}
{}
{}
\makeatother
\fi

\numberofauthors{3}
\author{
\alignauthor Michel Rottleuthner \\
    \affaddr{HAW Hamburg}\\
    \email{michel.rottleuthner@haw-hamburg.de}
\alignauthor Thomas C. Schmidt \\
    \affaddr{HAW Hamburg}\\
    \email{t.schmidt@haw-hamburg.de}
\alignauthor Matthias W\"ahlisch \\
	\affaddr{Freie Universit{\"a}t Berlin}\\
	\email{m.waehlisch@fu-berlin.de}
}

\title{Dynamic Clock Reconfiguration for the Constrained IoT\\ and its Application to Energy-efficient Networking\ifsqueeze\vspace{-0.1cm}\fi}

\begin{document}
\maketitle
\begin{acronym}[Bash]
	\acro{PLL}{phase-locked loop}
	\acrodefplural{PLL}[PLLs]{phase-locked loops}
	\acro{WSN}[WSN]{Wireless Sensor Network}
	\acrodefplural{WSN}[WSNs]{Wireless Sensor Networks}
	\acro{IoT}[IoT]{Internet of Things}
	\acro{MCU}[MCU]{microcontroller unit}
	\acrodefplural{MCU}[MCUs]{microcontroller units}
	\acro{IoT}[IoT]{Internet of Things}
	\acro{ADC}[ADC]{analog to digital converter}
	\acrodefplural{ADC}[ADC]{analog to digital converters}
	\acro{ALPME}[ALPME]{automatic low power mode entry}
	\acro{API}[API]{application programming interface}
	\acrodefplural{API}[APIs]{application programming interfaces}
	\acro{CCF}[CCF]{Common Clock Framework}
	\acro{COTS}[COTS]{commercial off-the-shelf}
	\acro{CPU}[CPU]{CPU}
	\acrodefplural{CPU}[CPUs]{CPUs}
	\acro{DFS}[DFS]{dynamic frequency scaling}
	\acro{DVS}[DVS]{dynamic voltage scaling}
	\acro{DVFS}[DVFS]{dynamic voltage and frequency scaling}
	\acro{FWSA}[FWSA]{flash wait state adaptation}
	\acro{ACG}[ACG]{automatic clock gating}
	\acro{LPM}[LPM]{low-power mode}
	\acrodefplural{LPM}[LPMs]{low-power modes}
	\acro{LUT}[LUT]{lookup table}
	\acrodefplural{LUT}[LUTs]{lookup tables}
	\acro{RAM}[RAM]{random-access memory}
	\acro{RC}[RC]{resistor-capacitor}
	\acro{RF}[RF]{radio frequency}
	\acro{RTC}[RTC]{real time clock}
	\acro{SPI}[SPI]{Serial Peripheral Interface}
	\acro{USB}[USB]{Universal Serial Bus}
	\acro{HWI}[HWI]{hardware independent}
	\acro{HWS}[HWS]{hardware specific}
	\acro{BASN}[BASN]{body area sensor network}
	\acrodefplural{BASN}[BASNs]{body area sensor networks}
	\acro{I2C}[I2C]{Inter-Integrated Circuit}
	\acro{OS}[OS]{operating system}
	\acro{PWM}[PWM]{Pulse-width Modulation}
	\acro{PU}[PU]{performance utilization}
	\acro{CMOS}[CMOS]{Complementary Metal-Oxide-Semiconductor}
	\acro{GPIO}[GPIO]{general-purpose input/output}
	\acro{HAL}[HAL]{hardware abstraction layer}
\end{acronym}

	\begin{abstract}
		Clock configuration takes a key role in tuning constrained general-purpose microcontrollers for performance, timing accuracy, and energy efficiency.
		Configuring the underlying clock tree, however, involves a large parameter space with complex dependencies and dynamic constraints.
		We argue for clock configuration as a generic operating system module that bridges the gap between highly configurable but complex embedded hardware and easy  application development.		
		In this paper, we propose a method and a runtime subsystem for dynamic clock reconfiguration on constrained \ac{IoT} devices named \sname.
	\sname derives measures to dynamically optimize clock configurations by abstracting  the hardware-specific clock trees. The \sname system service grants portable access to the optimization potential of dynamic clock scaling for applications. 				
	  We implement the approach on the popular IoT operating system RIOT for two target platforms of different manufacturers and evaluate its performance in static and dynamic scenarios on real devices.
		We demonstrate the potential of \sname by designing a platform-independent \ac{DVFS} mechanism that enables RIOT to autonomously adapt the hardware performance to requirements of the software currently executed.
		In a use case study, we manage to boost energy efficiency of constrained network communication by reducing the MCU consumption by \SI{40}{\percent} at negligible performance impact.
	\end{abstract}

	\category{D.4.9}{Operating Systems}{Systems Programs and Utilities}
	\category{B.8.2}{Performance and Reliability}{Performance Analysis and Design Aids}

\terms{Design, Management}

	\keywords{Embedded Systems, Energy, DVFS}

	\section{Introduction}
\label{sec-intro}

Embedded systems based on \acp{MCU} and diverse peripherals are omnipresent today, and the rapidly evolving \ac{IoT} deployment turns them into networked devices.
Hardware abstractions available from various (open source) \ac{IoT} operating systems~\cite{dgv-clfos-04, lcggp-mcse-17, bghkl-rosos-18} facilitate to assemble and maintain portable software on heterogeneous embedded platforms. Nevertheless, key performance characteristics  (\eg accuracy, energy, connectivity, lifetime) have conflicting impacts, which calls for advanced optimization and control of the interplay between hardware and software.

\begin{figure}
	\includegraphics[width=\linewidth]{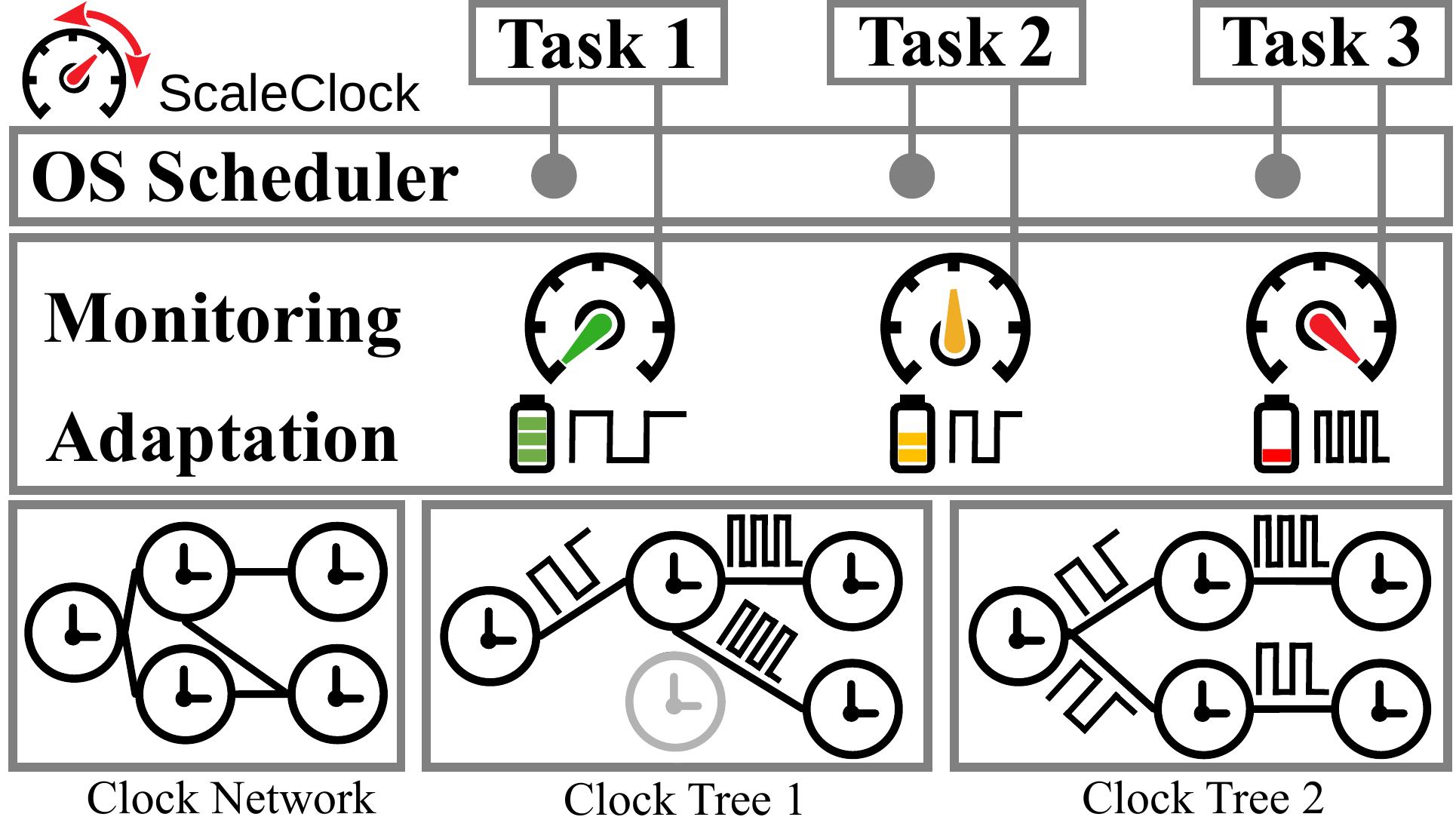}
	\caption{\sname derives tasks characteristics at runtime and adapts the system performance to application needs. Its clock tree abstraction it is able to dynamically reconfigure the hardware for more efficient operation and saving energy.}
	\label{solutionoverview}
	\postfigspace
\end{figure}

Energy availability is severely constrained on wireless \ac{IoT} devices and mandates for efficiency improvements on each layer~\cite{jtotd-aemws-07,rsw-sypea-21}.
The clock configuration of a system comprises a control knob for balancing between high performance and low power as every switched signal inevitably affects power consumption~\cite{ee-fduor-11}.
Adaptive software can exploit this to save energy (\eg via \acf{DVFS}), making the clock configuration fundamentally important whenever energy efficiency is a concern~\cite{cbba-pcmde-10}.

Frequency scaling requires access to the clock configuration of the underlying hardware. This should be supported by the hardware abstraction layer (HAL) of the operating system.
Current operating systems for embedded devices, though, lack abstractions for clock trees and hence are  unable to manage the full system clock configuration space. They cannot  dynamically manage clock configurations for optimized performance and energy efficiency~\cite{kbw-iruws-16, aasma-icdvf-20, cagll-pcdmc-21}. 
This wastes energy in particular on low-power, wireless IoT devices, where peripheral buses are often one order of magnitude slower than the \acsu{CPU} and common radio interfaces have even lower data rates (\eg IEEE802.15.4 with \SI{250}{\kilo\bit\per\second} and LoRa with $<$\SI{1}{\kilo\bit\per\second}).

In this paper, we introduce \sname, a lightweight generic clock configuration solution for constrained IoT devices.
\sname enables platform-independent control of complex, \ac{MCU}-internal clock settings during runtime.
\sname configures the \emph{clock network} of a device, the set of clock sources, modifiers, and consumers, such that the effective \emph{clock tree} reflects application needs (see \autoref{solutionoverview}). 

Dynamic power management with reconfiguration of clock trees on IoT devices is challenging for two~reasons.
First, constrained devices are restricted by processing power and memory.
This prevents transferring approaches from prior work such as Linux.
Second, most \ac{MCU} manufacturers provide a variety of internal and external clock sources.
These sources exhibit diverse properties  supporting orthogonal objectives.
We want to integrate those for the sake of a versatile, platform-independent IoT.

\sname overcomes these challenges by generic methods for assessing tasks and identifying their optimized clock configuration.
It utilizes a layered architecture consisting of composable base types for clocks and unified configuration interfaces that decouple from the underlying hardware.
Applications do not need to interact with our framework because the \sname transition manager autonomously interacts with the OS~scheduler and clock configuration.
\sname is implemented as part of the open-source operating system RIOT and validated on two common IoT platforms (STM32 and EFM32).
Despite its flexibility, the overhead remains small (\eg 4\% more memory than the platform-specific clock tree).
It optimizes energy during runtime and explores all possible clock configurations when needed.
This operation is fast and results are cached.
It provides \ac{DVFS} as a lean OS-centric service and saves more energy than the race-to-idle strategy often used on constrained devices.

In summary, our main contributions are:

\vspace{-2pt}

\begin{enumerate}
	\item A system service for clock-tree exploration and dynamic reconfiguration, which uses a novel method to proactively assess task characteristics for optimizing \ac{DVFS} control.
	(\autoref{sec-concept})
	\item A flexible abstraction for a light-weight cross-platform modeling of \ac{MCU} clock-trees. (\autoref{sec-impl})
\item Comprehensive evaluations, including the validation of \sname on two independent target platforms. (\autoref{sec-eval})
	\item A case study demonstrating how \sname improves energy efficiency of low-power communication. (\autoref{sec-networking})
	\item A \sname  open source implementation on RIOT.
\end{enumerate}
\vspace{2pt}

In the remainder of this paper, we introduce the problem space and potentials of a reusable model of clock trees with (re-)configurable clocks in \autoref{sec-ps}.
We present our core contributions in \autoref{sec-concept}--\autoref{sec-networking}, discuss related work in \autoref{sec-rel-work}, and summarize our findings and present an outlook in \autoref{sec-conc}.

 	\section{The Clock Tree and its Forest of Problems}
\label{sec-ps}
Clock trees comprise the low-level clock networks within \acp{MCU}. They are responsible for distributing and conditioning the clock ticks required for operating almost every internal component.
They substantially differ in complexity and implementation -- not only between vendors but also between \ac{MCU} series and models of the same vendor.
In addition, clock trees often come with very complex dependencies that are entangled with hardware configurations and constraints, use of peripherals, environmental conditions, and application demands.
Modeling the clock tree configuration as generic reusable component is therefore challenging.

\begin{figure}
	\includegraphics[width=\linewidth]{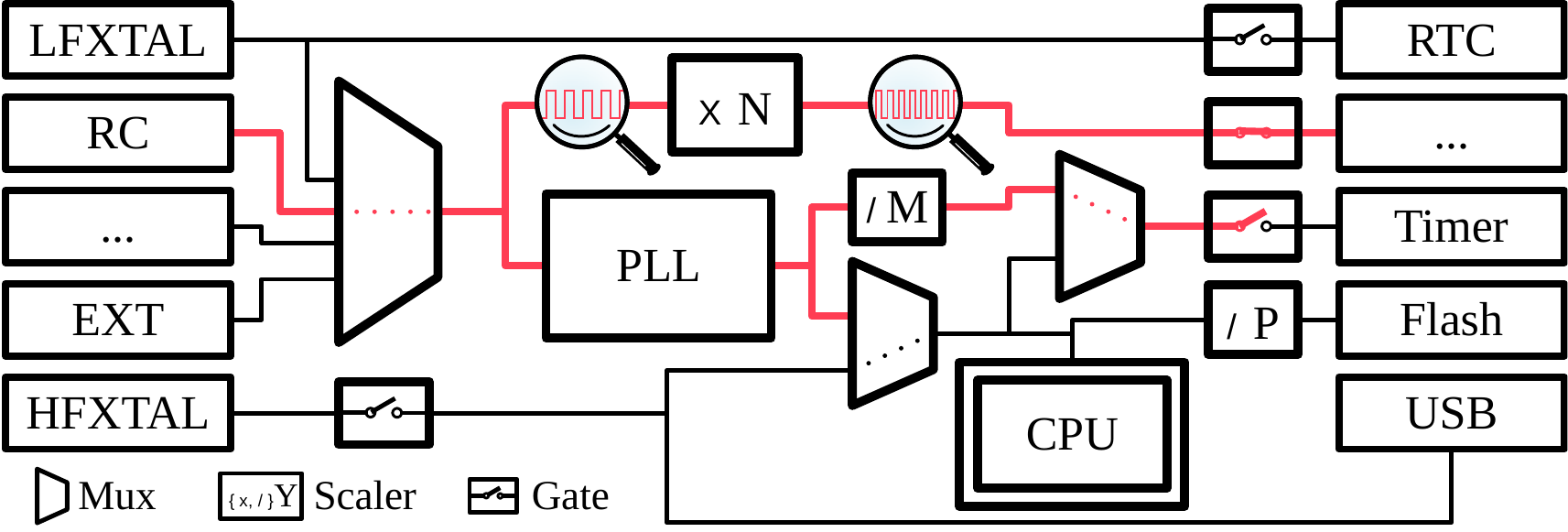}
	\caption{Example clock tree consisting of various clock sources (left), muxes, scalers, gates, and consumers (right).}
	\label{clocktree}
	\postfigspace
\end{figure}

\autoref{clocktree} illustrates a simplified example of an \ac{MCU} clock tree.
Left are various clock sources of the tree.
Sources provide clock signals to intermediate nodes such as gates, muxes, and scalers to eventually feed a consumer node (\eg a timer or the \acs{CPU}).
Intuitively it may seem appropriate to just put configuration logic of each part to respective peripheral drivers.
Drivers, however, only control local parts (\ie leaf nodes) and cannot manage the overall coordination of clocks that affect multiple devices. The latter  requires a global component on the kernel level.

We surveyed the clock configuration features of eleven popular embedded \ac{IoT} operating systems.
The majority (\ie Contiki, FreeRTOS, LiteOS, Mbed, Mynewt, NuttX, RIOT, TinyOS) avoid this problem by implementing the clock configuration in static code that is only configurable before compilation.
While being efficient w.r.t. code size and execution time, this static pattern does not allow for dynamic reconfiguration at runtime. It effectively neglects optimization potentials, since a single static configuration that operates in the energetic sweet-spot of every task and hardware does not exist.
Other OSs (\ie ChibiOS, Tock, Zephyr) only provide  partial dynamic features.
Specifically, ChibiOS offers an \ac{API} for switching between static platform specific presets by re-initializing the entire clock subsystem, but lacks  support for exploration, topology control, and performance assessment.
Tock provides an \ac{API} for clock gating (\ie enable and disable control), which can be used for automated peripheral power management.
Zephyr has an interface to access frequency properties of specific clocks.
However, none of them considers an abstract model of the clock tree, topology control, exploration of possible configurations, nor dynamic performance adaptation.

To the best of our knowledge, there is  currently  no viable solution for handling advanced clock \mbox{(re-)configuration} features even though many operating systems could profit from dynamic optimizations.

\subsection{Common Building Blocks for Clock Trees}
\label{sect-common-build-block}
Clock trees involve many properties that largely vary between devices.
Still, most  entities (\ie \textit{clock nodes}) that compose a clock tree share common functionality to serve  specific high-level purposes.
They either manipulate the topology of clock signals (\ie the routing to different destinations) or change properties of a clock, such as its frequency, calibration, or duty-cycle.
The most primitive form of a clock node is a \textit{gate}.
Its sole purpose is to enable or disable the forwarding of its input clock to its output.
A \textit{mux} selectively passes one of its multiple input clocks to its single output.
Multipliers (\eg based on \acp{PLL}) and dividers scale the frequency of the clock signal.
We therefore collectively refer to them as \textit{scalers}.

The exact features and implementations of those clock nodes differs between platforms but their high level objective and semantics are shared.
Respectively, the hardware interfaces admit many commonalities.
For example, settings are commonly programmed by writing values to specific memory mapped configuration registers, which also contain bit flags to trigger and observe operations and state changes.

\subsection{Configuration Space(s)}
Management and encoding of configuration parameters that describe the clock tree and its hardware properties face challenges.
Inherent features of hardware or its individual composition with external components must be visible to the configuration tools.
We differentiate between \textit{static properties}, \textit{static configuration} and \textit{runtime configuration}.

\paragraph{Static Properties}
Static properties are immutable for a specific target platform.
Here \emph{target platform} refers to a system model consisting of a particular \ac{MCU} and permanently connected components such as sensors and crystal oscillators (\eg a particular smartwatch or evaluation board).
Some properties are inherent to the \ac{MCU} model, \eg the possibility to connect an external crystal.
This optional oscillator may then be present on a specific target platform or not.
Overall, these static properties relate to different levels of abstraction and should be defined on the topmost level of commonality. This can be  
 \acp{MCU} with a specific instruction set,  an \ac{MCU} family, an \ac{MCU} model, or an actual  user device.

\paragraph{Static Configurations}
Static configurations concern parameters that are adjustable per se but their static implementation enforces a fixed value at runtime.
Adjusting those parameters therefore requires recompilation.
For example, a device that reads data from an external pen drive could be statically configured to enable a \SI{48}{\mega\hertz} clock for driving the \ac{MCU}-internal \ac{USB} peripheral.

\paragraph{Runtime Configurations}
Runtime configurations apply new system parameters during operation.
Data describing valid configuration values and adaptation logic must be part of the firmware.
For the above \ac{USB} example, runtime configuration of the peripheral clock becomes highly preferable in battery powered scenarios as it reduces power consumption while the \ac{USB}-port is unused.

These simple examples of different configuration types illustrate that there is no universal solution that serves all scenarios.
Though, a self-adaptive system that shall adjust to its current needs, requires runtime configuration.

\subsection{Complex Clock Configuration Transitions}
\label{subsection_complex_clock}
Procedures for changing the configuration of a clock (\ie its topology or frequency) can vary significantly in complexity.
This applies to the execution of the transition itself but also to side effects when other parts of the clock tree are affected by a change. 
In simple cases the clock may for example be updated on-the-fly by just writing a new pre-scaler value to the corresponding configuration register.
The change is immediately applied and a new frequency is in effect.
Contrarily, we refer to a  \textit{complex transition} if it consists of multiple phases.
There are several reasons why complex transitions are needed.
A Transition may affect parts further down in the clock tree, demanding for pre- or post-operations to make the new configuration applicable.
Examples are adjustments to voltage range or flash access parameters.
There are also changes which take time to complete and temporarily produce unstable frequencies, \eg on \acp{PLL} that require some time to stabilize.
Other clocks are  simply not adjustable on-the-fly.
This is a particular problem when users of that clock are at the same time uninterruptible.
Complex transitions solve this by a temporary switch to an alternative source, performing the adjustment, and then switching back.

Determining transitions and target configurations online foremost requires semantic encoding of what each involved step does in order to detect if they can be applied or raise conflicts.
Since this kind of online exploration can impose significant overhead, pre-calculating complex clock transitions is considered beneficial for fast repeated execution.

\subsection{Hard Resource Limits}
It is worth highlighting the significant challenges associated with scaling down the overhead of a generic solution to constrained \ac{IoT} devices.
In those small embedded systems data structures to encode hardware properties and runtime management of configurations can significantly impact the memory footprint.
Dynamic loading of hardware description files is out of scope and even dynamic memory allocation is preferably avoided for keeping a fixed memory budget~\cite{lmpsw-tossn-05}.

Overall, its complexity poses a severe challenge on the design of run-time methods to (re-)configure clock trees.
 	\section{\sname Approach for Self-Optimization}
\label{sec-concept}
This section introduces the core concepts of \sname and describes the mechanisms with which \sname enables the system to assess tasks and optimize their energy level. Built on a generalized clock subsystem control, \sname introduces the ability to dynamically adapt clock frequencies to execution demands and thereby leverage potential energy savings without sacrificing application demands.

\paragraph{Dynamic Power Consumption}
\label{dyn-power}
Essentially, power consumption of \ac{CMOS} circuits can be separated into a \emph{dynamic} and a \emph{static} part.
We focus on scenarios, in which an \ac{MCU} performs computations, and the dynamic share dominates the power consumption---as we will also show in \autoref{sec-eval}.
The dynamic power consumption of a switched circuit is described by \begin{math}P = \alpha \cdot C \cdot V^2 \cdot f\end{math}, where 
$C$ is the capacitance of the switched circuit, $V$ the voltage, and $f$ its frequency.
The transistor switching activity (\ie number of switched transistors) is reflected by $\alpha$.
$C$ is a property of the specific device, and the executing application mostly defines $\alpha$. Hence we can only influence  the parameters  $V$ and $f$ to reduce power consumption of a given application on a  device.
Both are subject to the following conceptual and practical limits.
\one The minimal voltage depends on the frequency; thus, voltage cannot be reduced independently.
\two The  available range of voltage adjustment is significantly confined. On modern microcontrollers, core voltage often ranges from \SI{1.8}{\volt} down to around \SI{1}{\volt}, whereas frequencies can be scaled from hundreds of \SI{}{\mega\hertz} down to \SI{}{\kilo\hertz}.
\three Time overhead can differ significantly for scaling voltage vs.\ frequency. Voltage scaling incurs a relatively static time overhead in the order of several \SI{}{\micro\second} per \SI{10}{\milli\volt}. Frequency scaling can be almost instantaneous (\eg when switching a mux or adapting a scale value) to taking multiple \SI{}{\milli\second} (\eg when cold-starting an oscillator). \four At very low frequencies the  static power consumption becomes more relevant, which reduces efficiency.
For a more comprehensive background reference, we refer the reader to Castagnetti \etal \cite{cbba-pcmde-10} and Eyerman \etal \cite{ee-fduor-11}.

\subsection{Resource Demands are Dynamic}
The potential for dynamic energy optimization  (\eg from \ac{DVFS}) is significant because applications rarely utilize the full performance provided by the \ac{MCU}.
A mismatch between the performance configuration of the hardware and the utilization by the software inevitably wastes energy.
Full computing performance  is not required, for example, at execution phases in which the \ac{MCU} waits for an operation to complete, such as reading an external sensor, erasing a flash page, or transmitting data. This leaves potential to trade underutilized performance for energy savings.

On resource constrained devices race-to-idle is commonly employed because of its easy implementation, but it often lowers energy efficiency compared to more adaptive methods~\cite{kih-rpite-15}.
With this in mind, the demand for dynamic performance adaptation as a generic system service becomes apparent.
Its largest challenge is to define a generic mechanism for dynamic clock configuration on constrained devices.

\subsection{Assessment of Resource Utilization}
\label{conc:pu_ctl_dvfs}

A versatile feedback loop for dynamic runtime optimization needs to acquire precise knowledge about the system conditions via a simple, yet expressive metric. Consider a time slice $t$, of which the scheduled task utilizes the CPU for the fraction $t_{busy}$, then
\vspace{-0.3cm}
\begin{equation}
\label{eq:simple_util}
Load = \frac{t_{busy}}{t_{busy} + t_{idle}}
\end{equation}
defines a simple utilization metric often employed on \acp{MCU}.
We argue that this metric is not well suited for dynamic performance control because it is insensitive to tasks which only \emph{appear} to utilize the \ac{CPU} but in reality are limited by other operations.
Instead, we propose  a utilization metric that compares the actual utilization at two different frequencies $F_1$ and $F_2$.
For any scheduled time slice $t$, \autoref{eq:perf_util} relates the ratio between busy times at different frequencies to the frequency change ratio.
\begin{equation}
\label{eq:perf_util}
PU = \frac{t_{busy}(F_{1})}{t_{busy}(F_{2})} \cdot \frac{F_{1}}{F_{2}},\quad F_{1} < F_{2}
\end{equation}
For perfectly scalable tasks the busy time reduces by the same ratio as the frequency increases.
On the converse, tasks with low \ac{PU} values scale worse but show high potential for energy savings.
Notably, this definition drops the explicit use of the idle time, because the ability of a task to (not) scale well with frequency is independent of its idle time---in contrast to the global system load.
Idle time is considered implicitly as the time allocated to the idle task.

\subsection{Dynamic Scaling}

In \sname, we devise an online \ac{PU} assessment that executes on the device itself.
It instruments the operating system to collect the context switching count together with busy- and idle times for each task.
The core frequency is then opportunistically adapted while collecting measurement points for the \ac{PU} metric.
Based on these measurements a task-specific, energy-optimized frequency is selected.
For a given set of possible core frequencies corresponding clock tree configurations are determined in an exploration phase, which runs once on system init and on topology changes.

The dynamically assessed target frequency is set up before scheduling the next task.
Core voltage and flash wait state adaptation follow the frequency selection according to static constraints which encode clock-node and frequency-specific hardware limits.
A policy setting governs whether fast flash or low voltage is preferred in cases where they mutually exclude each other.
We evaluate this concept in \autoref{sec-eval} for the RIOT \cite{bghkl-rosos-18} operating system on real hardware.

\subsection{Interfacing Functional Capabilities}
\label{sec-conc-impl-capabilities}
Clock types implement different sets of capabilities, which we reflect by an individual assignment.
A clock can for example be \textit{scalable}, \textit{muxable}, \textit{gateable}.
Each capability provides its own interface functions such as accessing and configuring a scaling factor for a \textit{scalable} clock, configuring the parent for a \textit{muxable} clock, or simply enabling or disabling a \textit{gateable} clock.
Drivers that only differ in details  can reuse most capability code by only replacing the selective parts of the interfaces that differs.
Capability implementations are explicitly separated from the mapping functions that translate register content to logical values in order to reuse code sections wherever possible.

All properties described so far are provisioned as separate static data structures to support static memory allocation and selective inclusion.
These static definitions also open the door for \textit{a priori} encoding tweaks towards our design goals to minimize memory consumption and maximize execution speed.
Combining these principles allows an easy to understand modeling while still being expressive, memory efficient, and flexible to adjust for optimizations.
 	\section{\sname Implementation}
\label{sec-impl}

A suitable level of abstraction is essential for a reusable design that hides hardware specifics but grants sufficient access.
In \shortautoref{sect-common-build-block}, we identified a significant part of common behavior at the level of individual base elements (\ie \textit{gate}, \textit{mux}, and \textit{scaler}).
The hardware facing part of the interface sits at this level so that higher order clock trees can be flexibly orchestrated from separate clock instances.

Albeit uncommon clock nodes exist (\eg \acp{PLL} with multiple stages and outputs), even among devices with very complex clock trees we never encountered any that could not be modeled by combining the aforementioned base elements.
Moreover, custom clock types can still be added, and the design is widely applicable to \acp{MCU} with memory mapped configuration registers.
We thus argue that these clock types form a reasonably reduced but expressive set to model all typical \ac{MCU} clock trees.
We will demonstrate its utility later in \autoref{sec-eval}.
Key design decisions are presented next, for further details we refer to our publicly available code (see  \autoref{sec-conc}).

\subsection{Layered Architecture for Flexibility}
\label{sec-conc-impl-arch}

Our layered architecture puts common functionality and patterns to a higher utility layer and strictly separates data and code.
We observe that clock configuration can be done by adapting \textit{frequency} or changing \textit{topology}. 
Topology operations may affect the frequency whereas frequency operations shall never affect the topology, as this may have significant impact (\eg on clock availability and accuracy).
We explicitly separate this functionality to retain direct control.

\begin{figure*}[t]
\hfill
\begin{minipage}[t]{.32\textwidth}
\begin{lstlisting}[label=listing-node-read,caption={Get clock properties.},captionpos=b,escapeinside=||]
|\includegraphics{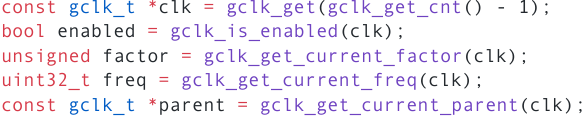}|
\end{lstlisting}
\end{minipage}
\postfigspace
\hfill
\begin{minipage}[t]{.26\textwidth}
\begin{lstlisting}[label=listing-node-write,caption={Set clock properties.},captionpos=b,escapeinside=||]
|\includegraphics{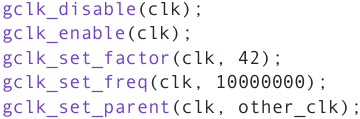}|
\end{lstlisting}
\end{minipage}
\postfigspace
\hfill
\begin{minipage}[t]{.35\textwidth}
\begin{lstlisting}[label=listing-dfs-cycle,caption={Set different core frequencies.},captionpos=b,escapeinside=||]
|\includegraphics{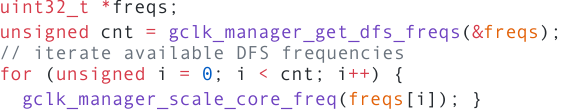}||\vspace{0.09cm}|
\end{lstlisting}
\end{minipage}
\postfigspace
\hfill
\end{figure*}

\begin{figure}[t]
	\includegraphics[width=\linewidth]{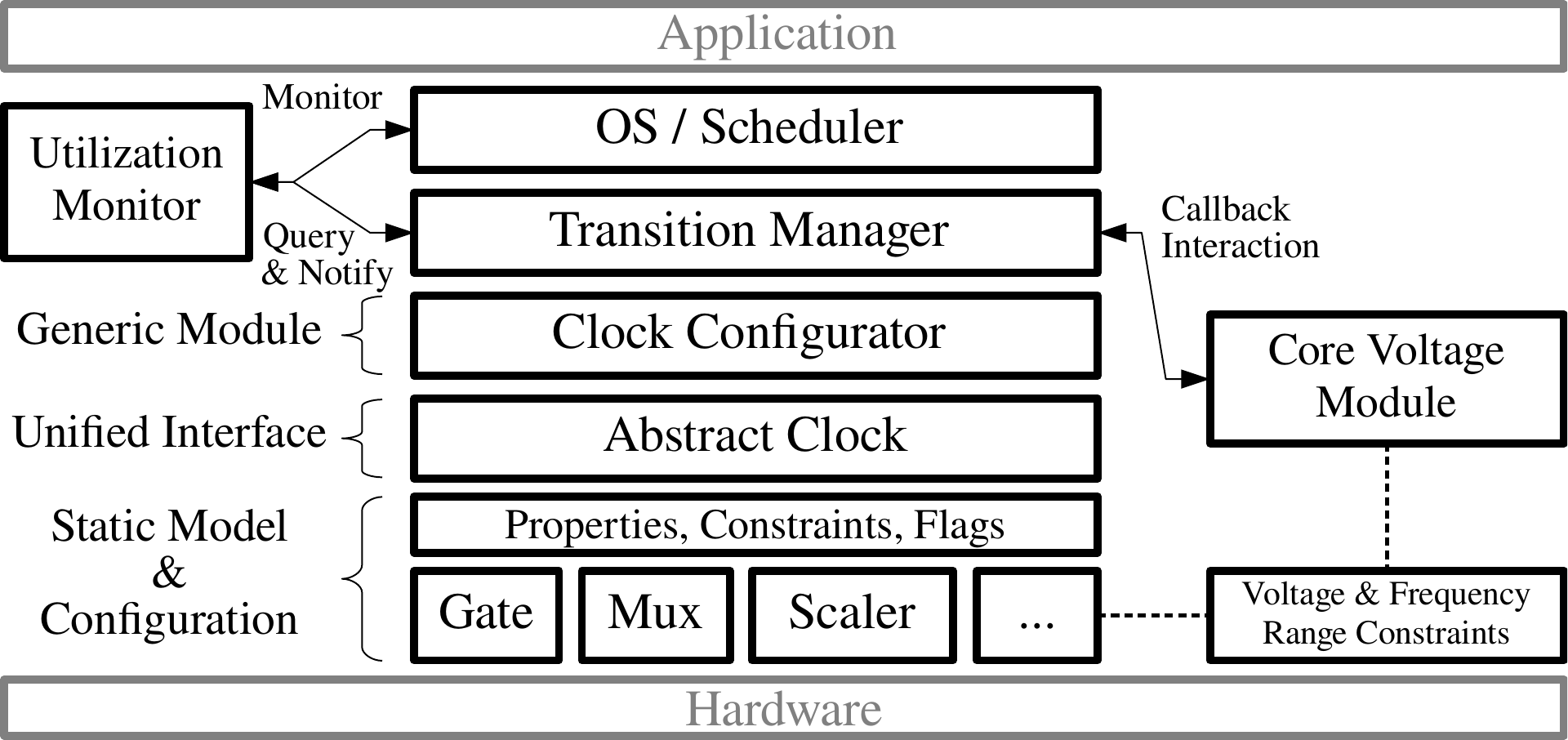}
	\caption{The \sname architecure incorporates a unified configuration interface to individual clocks. The transition manager uses OS statistics to self-adapt the system for low energy or high performance operation.}
	\label{fig-architecture}
	\postfigspace
\end{figure}

\autoref{fig-architecture} gives an overview of the \sname architecture.
An application never has to interact with the clock framework directly, unless it aims for manual control.
Operations on the \textit{Abstract Clock} interface are decoupled from the underlying hardware.
Platform specific configuration options and hardware access description are provided as static data, separated from the base clock functions, which are defined as independent driver functions.
Individual mapping functions  convert between semantic values (\ie multiplier, enable state, clock instance, \etc) and configuration register values.
Clock base types (\textit{Gate}, \textit{Mux}, and \textit{Scaler}) are provided for reuse by platform specific implementations.
Custom clocks can be modeled by separately orchestrating and decorating primitive operations, value descriptors, and mapping functions.
Tree variants are constructed from a list of parent choices of each clock. Pointers to unique static descriptors of clock instances serve as identifiers.
For easy traversal, getting the current parent of a clock returns a pointer to its static descriptor (see \autoref{listing-node-read}).
A topology entry datatype stores the logical configuration state of a clock instance for virtual representation of topology settings.

The clock configurator can query and reconfigure hardware state, \eg to get the current frequency or set a new scaling factor (see \autoref{listing-node-read} and \autoref{listing-node-write}).
The transition manager handles high-level tasks such as core frequency updates (\autoref{listing-dfs-cycle}), complex transitions as introduced in \shortautoref{subsection_complex_clock} (\autoref{listing-topo-transition-cycle}), and triggering notifications on configuration changes.
The utilization monitor gathers scheduler statistics (idle-, execution-time, context switches) to meter the performance of running threads and provides feedback to identify bottle-necked operations for dynamic optimization.
Energy is then saved by aligning the system performance to the task demand.
\autoref{listing-pua} shows a manual invocation of the \ac{PU}-assessment mechanism in a multi-threaded application.
\begin{lstlisting}[label=listing-topo-transition-cycle,caption={Code excerpt to switch between core topologies.},captionpos=b,escapeinside=||]
|\includegraphics{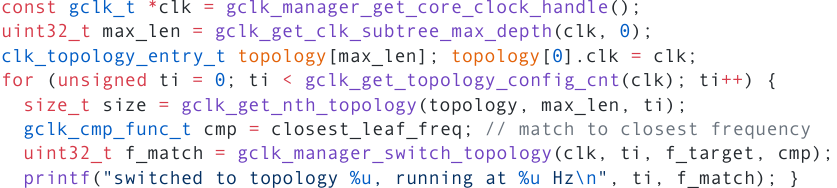}|
\end{lstlisting}
\vspace{-0.23cm}
\begin{lstlisting}[label=listing-pua,caption={Code excerpt to assess performance utilization.},captionpos=b,escapeinside=||]
|\includegraphics{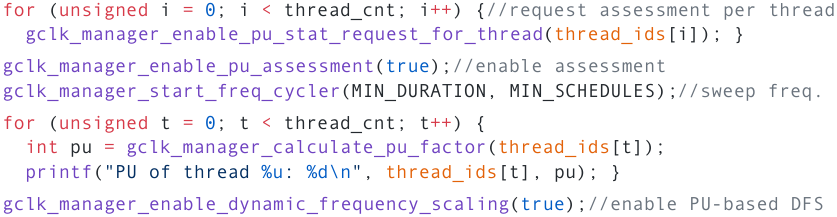}|
\end{lstlisting}

Porting \sname to new platforms only requires provisioning of hardware-specific static data (register mappings, constraints etc.), \ie layers below the abstract clock interface in \autoref{fig-architecture}.
Generic low-level drivers can be reused directly in most cases and manual effort to extend capability drivers of specific clocks (as described in \autoref{sec-conc-impl-capabilities}) is rarely needed.
Higher layers of \sname require no porting.

\subsection{Property Encoding}
Clock nodes can have many (configurable or immutable) properties, such as valid input sources or divider values.
Those must be accessible to upper layers, whereas respective register values and their addresses must be available to the hardware facing layer.
Reconfiguration constraints (\eg on-the-fly adaptability of scaling factors) must be stored too.

Numerical values are encoded using an extendable set of mapping types to ease modeling of sets of distinct values or ranges with associated semantics.
Memory overhead is further reduced by using indicators for common implicit value encoding patterns and modifiers.
In many cases this halves the storage for mappings between logical and register values.

Instead of saving configuration registers and masks separately for every clock node we leverage that only a subset of registers are typically responsible for clock configuration and combine them into a lookup table.
A register is then identified with only a few bits and a single integer can be shared for storing the register id, masking information and bit indexes for common flags (\eg enable or ready states).
Respectively,  a frequency multiplier descriptor that accepts factors between 1 and 8 by writing values from 0 to 7 into a specific register field can be encoded by the bounds (1,8) and a \textit{zero-based} implicit register value modifier.
Using the condensed configuration register descriptor, only two 32-bit integer values are needed to encode how to access the hardware register and which values it may attain.

\subsection{Notification \& Transaction Mechanism}
\label{sec-conc-impl-notification}
Interaction between \sname and other modules requires methods to request, indicate or block changes.
While \sname  itself can handle dependencies and constraints internal to the clock tree, this does not cover dependencies on internal state of peripheral drivers or application logic.
A clock, for instance, can be safely turned off if no other component is using it.
If it is used, \eg by a peripheral driver, it can block or allow transitions depending on its operation.

External modules can prepare for clock transitions or trigger reconfiguration procedures afterwards via \emph{pre}- and \emph{post}-hooks, which can be registered for any existing clock instance.
Shared peripherals can lock access for the duration of the transition via the \emph{pre}-call.
During the \emph{post} hook, the respective module must be put back to normal operation mode.

 	\section{Evaluation of \sname}
\label{sec-eval}
We are now ready to evaluate key performance metrics of \sname regarding \emph{functionality}, \emph{performance}, and \emph{overhead}.
First, we  describe the core features enabled by \sname and compare them to alternative mechanisms.
Second, we benchmark the effect of configuration parameters in static scenarios (\ie running different tasks at preset frequencies).
Third, we evaluate the performance of dynamically applying \sname  (\ie changing frequency at varying application needs). We measure the energy savings and the temporal overhead. 
Later in \autoref{sec-networking} we will assess the energy savings in a realistic case study of low-power wireless networking.

Experiments are conducted on the Nucleo-L476RG by STMicroelectronics and the SLSTK3402A EFM32 Pearl Gecko PG12 board by Silicon Labs.
Both boards are unmodified and our experimentation firmware sources are publicly available (see \autoref{sec-conc}) to ease reproducibility.
All stated current values relate to a static supply of \SI{3.3}{\volt}.
We use a highly accurate Keithley DMM7510 digital sampling multimeter~\cite{k-mdgsm-16}, connected to the power headers (IDD and BAT respectively).

\subsection{\sname Core Functions}
\sname supports active exploration of the clock subsystem and increases visibility and usability of hardware capabilities. Developers can use the same unified \ac{API} on all target devices to explore available clock configurations  and how they can be operated---instead of studying data sheets for each target platform.
The system itself exploits this knowledge to dynamically optimize the hardware configuration in concordance with the changing application needs.

\subsubsection{Reducing Power Consumption}
\label{subsubsect:pwr_con_red}
Applications facing (temporary) constraints on peak power consumption can instruct \sname to throttle clock speed.
For example, systems facing critical supply voltage may thereby limit power consumption to reliably maintain  operation of the voltage regulator.
Energy harvesting systems with varying power supply~\cite{sk-ehsns-11, basm-ehwts-16} can use this to significantly extend the runtime~\cite{aasma-icdvf-20}.
Power consumption can be reduced further by gating (\ie disabling) unused clock sources, sub-topologies of the clock tree, or input clocks of inactive peripherals.
Active power management can apply this dynamically~\cite{khzs-pmehs-07}.

We determine the magnitude of enabled power reductions  by executing several micro-benchmarks on our test systems.
\begin{figure}\centering
	\includegraphics{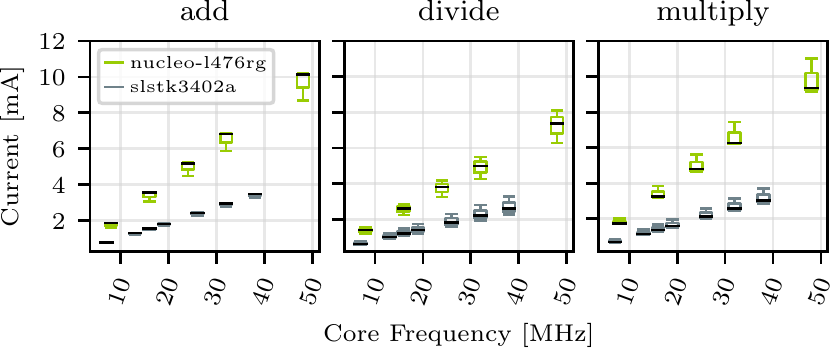}
	\caption{\ac{MCU} current draw behavior for tasks that execute loops of different integer instructions (add, divide, multiply) at varied clock frequencies on two devices. IQR: 25th-75th percentile, whiskers: Q1-1.5*IQR and Q3+1.5*IQR.
	}
	\label{fig_DFSI}
	\postfigspace
\end{figure}
\autoref{fig_DFSI} shows the current drawn at different \ac{CPU} frequencies set up by \sname.
We use the most compatible topologies in this experiment, in which the system is clocked by a scalable \ac{RC} oscillator.
Three aspects stand out. \one a roughly linear relationship between frequency and power consumption in all configurations; \two both platforms show significantly lower consumption for division. This can be traced back to divisions requiring more cycles, which reduces the proportion of other \ac{CPU} activities (\ie fetch, decode, memory access), lowering the switching activity; \three the ratio between current and frequency (\ie the slope) is significantly different on both targets.

\begin{figure}
	\centering
	\includegraphics{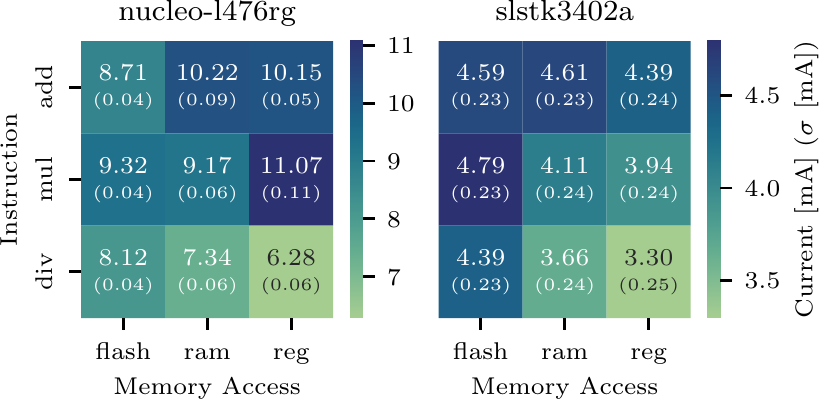}
	\caption{\ac{MCU} current draw behavior of two devices when executing tasks which loop different integer instructions (divide, add, multiply), using data from distinct memory locations (flash, RAM, registers); left: \SI{48}{\mega\hertz}, right: \SI{40}{\mega\hertz}. }
	\label{fig_TCI}
	\postfigspace
\end{figure}

Beyond  instruction execution, the consumption is also affected by the memory access required to fetch data.
\autoref{fig_TCI} shows this variation of power consumption when the same instructions use data from different memory locations.
This effect cannot be generalized into simple rules, as it has diverse impact, \eg for the \emph{nucleo-l476rg} using data from registers reduces consumption for division while increasing it for multiplication.
Both devices show a notable difference in how flash access affects the consumption.
The generally more efficient \emph{slstk3402a} exposes a much higher relative consumption when flash access is involved.
Overall, these results highlight two relevant insights.
First, operations of the executed task change the total consumption in a considerable magnitude.
Second,  precise knowledge of task behavior and device characteristics provide valuable runtime information for deciding whether a specific frequency configuration is energetically more efficient for that task.

\ac{MCU} power consumption is commonly reduced with duty cycling via \acp{LPM} in which any execution is stopped completely.
This binary on-off operation mode lacks gradual control over performance and consumption during active operation.
The \sname dynamic frequency adaptation can be used in complement to achieve additional energy savings during the active periods, in which duty cycling has no effect, and it does not interfere with duty cycling applicability.
\autoref{tbl_lpm_timing_overhead} puts our previous consumption statistics of a slowed down (but still actively processing) \ac{MCU} into perspective with the static consumption of low power modes, which completely stop ongoing execution.
It also lists the time overhead incurred when transitioning from the sleeping \ac{LPM} state back to operation mode as measured via \ac{GPIO} instrumentation.

\newcommand{\sml}{\footnotesize}
\newcommand{\SML}{\footnotesize}
\begin{table}\begin{center}
\caption{Time overhead for transitioning from a static low power mode to executing a function.}
\label{tbl_lpm_timing_overhead}
\begin{tabularx}{\columnwidth}{l l l r}
\toprule
\SML\textbf{Board} & \SML\textbf{LPM}     & \SML\textbf{Exit Transition Time}  & \SML\textbf{Current}  \\
\midrule
\sml nucleo-l476rg & \sml PM0 & \sml \SI{9.3}{\milli\second} (to main)        & \sml \SI{410.3}{\nano\ampere} \\
\sml nucleo-l476rg & \sml PM0 & \sml \SI{2.4}{\milli\second} (to system init) & \sml \SI{410.3}{\nano\ampere} \\
\sml nucleo-l476rg & \sml PM1 & \sml \SI{9.6}{\micro\second} (to callback)    & \sml \SI{8.4}{\micro\ampere} \\
\sml slstk3402a    & \sml PM0 & \sml \SI{18.9}{\micro\second} (to callback)   & \sml \SI{7.4}{\micro\ampere} \\
\sml slstk3402a    & \sml PM1 & \sml \SI{18.8}{\micro\second} (to callback)   & \sml \SI{7.8}{\micro\ampere} \\
\bottomrule
\end{tabularx}
\end{center}
\end{table}

\subsubsection{Improving Energy Efficiency}
Applications that want to maximize energy efficiency can use {\sname} to adaptively optimize the clock frequency at runtime.
There is potential to dynamically save energy during execution as long as the \ac{CPU} capacity is not fully utilized as noted in \shortautoref{conc:pu_ctl_dvfs}.
Tasks dominated by instructions that require access to \ac{CPU} registers or \ac{RAM} scale well with frequency and are often most efficiently executed at the highest applicable frequency because minimizing execution time reduces static loss related to the active \ac{MCU}~\cite{aasma-icdvf-20}.
Contrary, tasks slowed by I/O access or other asynchronous interactions execute more efficiently at a lower core frequency~\cite{cagll-pcdmc-21}.
Losses due to dynamic switching of instructions without progressing a task are avoided in those cases, which quickly outweighs static losses.
In practice, taking advantage of this unused potential requires either \textit{a priori} knowledge about tasks or some assessment at runtime.
\sname follows the latter variant, the benefit of which we quantify in the following.

\subsubsection{Impact of Dynamic Performance and Topology Control}
Our analysis up to this point indicates that the energy saving potential depends on task characteristics.
To capture other conditions which affect dynamic optimizations we measure how the energy efficiency of a given workload is affected by different topologies, core frequencies, and scaling methods.

\begin{figure}\centering
	\includegraphics{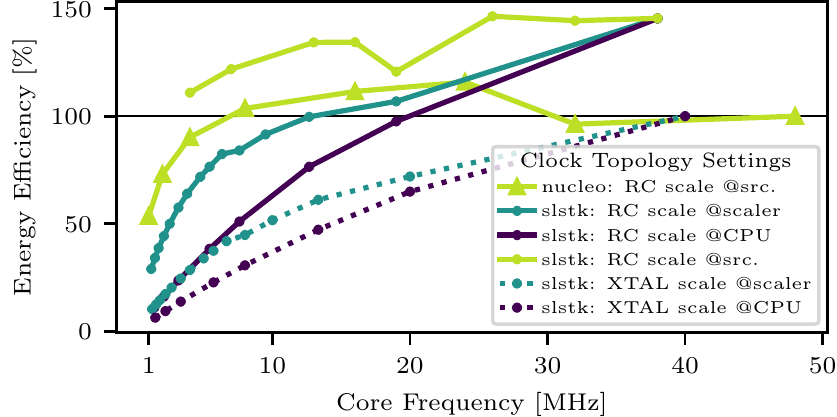}
	\caption{Energy efficiency while executing the same task at different topology and frequency configurations. Names refer to the clock source and the point in the topology where the frequency is scaled down. The \SI{100}{\percent} baseline marks the highest frequency of the default topology.}
	\label{fig_DFSI_speed_efficiency}
	\postfigspace
\end{figure}

The results in \autoref{fig_DFSI_speed_efficiency} relate the converged average energy efficiency and core frequency for different topology and scaling variants.
Displayed variants differ in type of clock source (RC vs XTAL) and the point within the topology path where the frequency adaptation happens (\ie as close as possible to either the source or CPU, or at an intermediate scaler).
The workload includes all permutations of instructions and memory access variants used before ($\textit{\{add, mul, div\}} \times \textit{\{reg, ram, flash\}}$, see \shortautoref{subsubsect:pwr_con_red}).
These operations are generally efficient at high frequencies as they are executed without busy waiting or asynchronous I/O. Therefore, the optimization potential for \ac{DVFS} is low---aside from  internal flash access that cannot keep up with the \ac{CPU} speed.

We chose the \SI{100}{\percent} baseline as the energy consumed at the highest frequency operation of \one the \emph{RC@src.} topology for the \emph{nucleo-l476rg} or \two the \emph{XTAL@scaler} topology for the \emph{slstk3402a}.
All configurations exhibit low efficiency at very low frequencies.
The largest efficiency impact (more than \SI{40}{\percent}) is obtained by selecting an \ac{RC} source over a crystal oscillator  (\emph{XTAL} vs. \emph{RC}).
The noticeable drop at \SI{20}{\mega\hertz} (\emph{RC@src} on \emph{slstk3402a}) follows from constraints of this particular system configuration which lead to an unfavorable ratio between flash speed and core frequency.
Another notable effect shows the \emph{nucleo-l476rg} device, at which the benefits of voltage scaling outweigh the efficiency penalty of lower frequencies between 8 and \SI{24}{\mega\hertz} giving more than \SI{15}{\percent} better energy efficiency.
Comparing configurations with the same source and frequency (\eg \emph{XTAL@x} or \emph{RC@x}) shows that energy efficiency improves with scaling a node that is closer to the clock source. 
We can conclude that already without dynamic optimizations \sname can improve energy efficiency by about \SI{15}{\percent} with scaling down voltage and frequency and over \SI{40}{\percent} by switching the source topology.

Recalling the linear relation between frequency and current (see \shortautoref{subsubsect:pwr_con_red}), we now question how the potential to reduce power consumption can be translated into further energy savings for applications that  do not require full \ac{CPU} performance.
We create a synthetic benchmark to explore the space of achievable energy savings.
It consists of 100 tasks with varying portions of computational and time-dependent operations.
We parameterize the task set from fully scalable computational workloads to purely time-dependent operations in order to cover the full range of possible workload characteristics between the extremes of scalability.
We measure the energy consumption for each task at each frequency and the energy-optimal setting is identified per task.
Regarding the use of synthetic benchmarks we note that the energy-related characteristics of a task are its execution time and its switching activity (\ie $\alpha$ as given in \autoref{dyn-power}). Varying those synthetically in a micro-benchmark reveals characteristics, but also transfers to other tasks as their timing is known (measured via \ac{PU} assessment), $\alpha$ is task-specific, and the \ac{PU} values are established per task.
An exception is that flash wait-state adaptation may slightly affect $\alpha$ towards lower frequencies (by reducing the wasted cycles for bottle-necked flash access). Yet, this effect can be fully isolated by performing the \ac{PU}-assessment without flash adaptation.

\begin{figure}[t]
	\centering
	\includegraphics{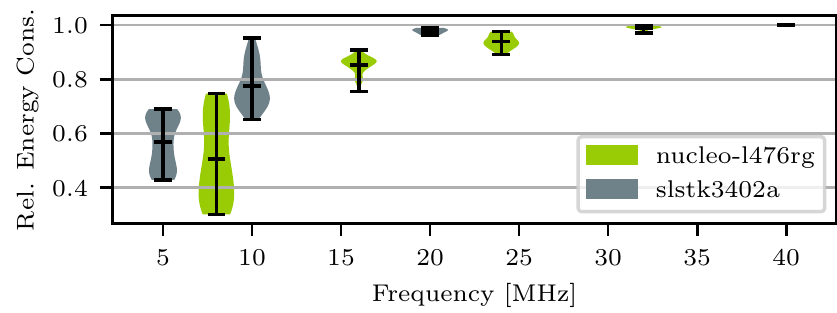}
\caption{Energy distributions for task sets executed at their energy-optimal core frequency, relative to  consumption at  highest freq. (\SI{40}{\mega\hertz}). Ticks: extrema and mean values.}
	\label{fig_DFSTI_energy}
	\postfigspace
\end{figure}

\autoref{fig_DFSTI_energy} displays the distribution of energy consumptions for all task subsets at their energy-optimal frequencies. 
Energy savings of more than \SI{60}{\percent} can be achieved if a task is most efficiently executed at a low core frequency (here \SI{8}{\mega\hertz}).
It is worth noting that only \ac{DFS} is applied in this case, of which we previously saw a reduced efficiency if applied to computation-intensive tasks.
Voltage scaling is expected to improve these further.

\subsubsection{Assessing Performance Utilization}
\begin{figure}[t]
	\centering
	\includegraphics{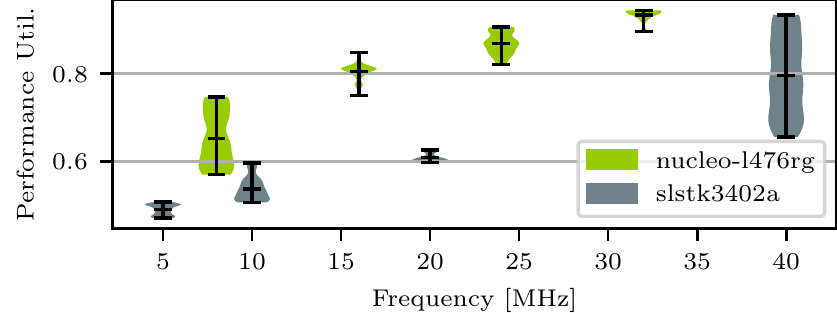}
	\caption{PU value distributions for the task sets executed at their energy-optimal core frequency. Ticks mark extrema and mean values.}
	\label{fig_DFSTI}
	\postfigspace
\end{figure}

 We now evaluate the \sname mechanism for dynamically identifying the optimal frequency, \ie the \ac{PU} metric for assessing the scaling potential of a task (see  \shortautoref{eq:perf_util}). 
\autoref{fig_DFSTI} shows the distributions of PU values for the same task sets as a function of their optimal frequencies. Per platform we observe a strictly monotonic relation between the optimal frequency and the PU values. This leads us to the conclusion that it is viable to deduce energy-optimal frequencies for the tasks from their respective \ac{PU} values. Hence, the metric justified its effectiveness for dynamically identifying the optimal frequency.

\begin{figure}\centering
\includegraphics{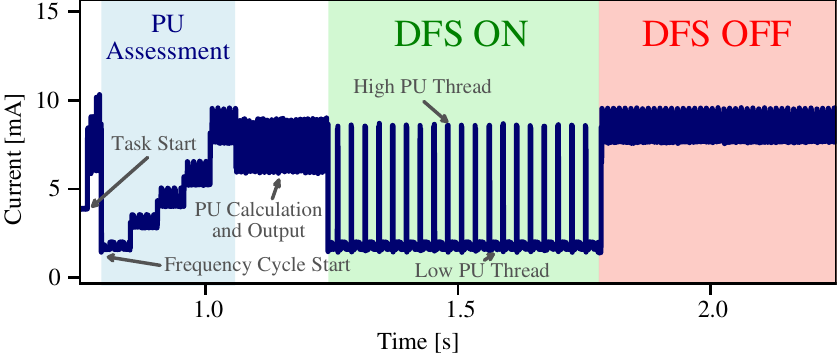}
\caption{\ac{MCU} current draw when performing online performance utilization assessment, comparing automatic \ac{DFS} with static high frequency operation.}
\label{fig_AATPU_hl}
\postfigspace
\end{figure}

Stirring the \ac{DVFS} functions from online assessments enables the system to automatically align performance with task demands.
\autoref{fig_AATPU_hl} shows the current profile of the \emph{nucleo-l476rg} \ac{MCU} while executing two threads that perform workloads of different kind.
Both threads start back-to-back and are triggered in an alternating pattern---a typical producer-consumer scenario.
One benefits from a higher core frequency (\ie high \ac{PU}, processing), while the other is most efficiently executed at a lower frequency (\ie low \ac{PU}, acquisition).
After the task is started, a stair pattern is visible related to the execution of the automatic \ac{PU}-assessment that cycles through multiple available frequencies while both threads still execute.
It is worth noting that the method introduced in \autoref{conc:pu_ctl_dvfs} is used with more than two frequencies here and the collected data of each frequency pair is used for the calculation of the \ac{PU} value.
Then the respective \ac{PU} factors are derived for both threads and written out together with debug information via a serial connection.
The duration of this step is governed by the serial communication while the overhead of the calculation is negligible.
Thereafter, the \ac{DFS} mechanism is enabled and reduces power consumption of the low \ac{PU} thread by more than \SI{70}{\percent}.
Even though the lower frequency of the low \ac{PU} thread also reduces the performance, the overall energy consumed for the same work is reduced by almost~\SI{40}{\percent}.

\subsection{\sname overhead}
The overhead induced by \sname is important for its utility.
We measure overhead in terms of memory and runtime of the base operations.
Base operations refer to initialization steps such as configuration exploration, as well as adaptation steps to change frequency and topology.

\subsubsection{Memory}
We differentiate between platform-agnostic parts and (static) data that encodes the platform-specific clock tree model.
Memory of the generic parts such as the clock configurator or transition manager is dominated by instructions whereas the clock tree model mainly consists of static data.
\autoref{tbl_AI} lists the memory overhead of the different \sname components.
Overall, the generic part of the \sname module uses $\approx$\,\SI{5}{\percent} of the total memory required by the test firmware used for this paper, whereas the platform specific clock tree model requires below \SI{1}{\percent} of the memory.

\newcommand{\na}{-}
\begin{table}
\caption{Memory used by the \sname building blocks.}
\label{tbl_AI}
\resizebox{\columnwidth}{!}{\begin{tabularx}{\columnwidth}{ l c c}
 \toprule \multicolumn{1}{l}{\SML \textbf{Component}} & \multicolumn{1}{c}{\SML \textbf{\texttt{ROM Size}}} & \multicolumn{1}{c}{\SML \textbf{\texttt{RAM Size}}}\\
 \multicolumn{1}{l}{\SML Generic clock}   & \multicolumn{1}{c}{\sml24 bytes} & \na \\
 \multicolumn{1}{l}{\SML Config register descriptor}   & \multicolumn{1}{c}{\sml32 bits} & \na \\
 \multicolumn{1}{l}{\SML Shared register \ac{LUT} entry} &  \multicolumn{1}{c}{\sml32 bits (single pointer)} & \na \\
 \multicolumn{1}{l}{\SML Zero based mux option} & \multicolumn{1}{c}{\sml32 bits (single pointer)} & \na \\
 \multicolumn{1}{l}{\SML \ac{LUT} mux option} & \multicolumn{1}{c}{\sml64 bits (pointer + reg. value)} & \na \\
 \midrule
 \multicolumn{1}{l}{\SML Clock configurator} & \multicolumn{1}{c}{\sml~~\SI{5}{\kilo\byte}} & \na \\
 \multicolumn{1}{l}{\SML Clock manager} & \multicolumn{1}{c}{\sml\SI{7.5}{\kilo\byte}} & \multicolumn{1}{c}{\sml\SI{172}{bytes}} \\
 \multicolumn{1}{l}{\SML Static clock tree model} & \multicolumn{1}{c}{\sml\SI{2.5}{\kilo\byte}} & \na \\\multicolumn{1}{l}{\SML Task PU data (per thread)} & \na & \multicolumn{1}{c}{\sml\SI{32}{bytes}} \\
 \bottomrule
\end{tabularx}
}
\end{table}

\subsubsection{Clock Tree Exploration}
Proper clock tree configuration usually forces developers to carefully study hardware data sheets~\cite{szs-amctc-16}.
{\sname}   substantially eases this duty by providing a unified interface for exploring, configuring, and testing clock configurations interactively.
On the \texttt{slstk3402a} platform all possible configurations for driving the core clock can be evaluated in less than three seconds. The faster \texttt{nucleo-l476rg} finishes this exploration in less than \SI{500}{\milli\second}.
Nevertheless, the overhead of exploring configurations is considered non-critical because this rarely needs execution  (\eg at initial boot) and the results can easily be cached.
Albeit \emph{a priori} exploration is preferable, \sname is fully capable of performing this step on demand.

\begin{figure}
	\centering
	\includegraphics{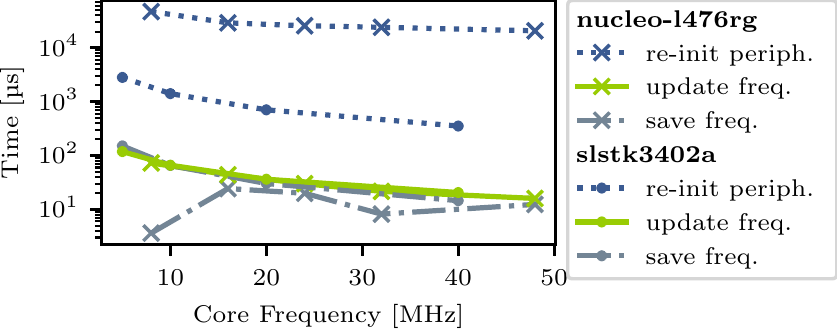}
	\caption{Timing overhead for simple frequency transitions performed at different clock frequencies.}
	\label{fig_ACT}
	\postfigspace
\end{figure}

\subsubsection{Time to Change Frequency}
Switching between different frequencies is more critical because it is expected to run frequently, \eg when employing \ac{PU}-based \ac{DVFS}.
\autoref{fig_ACT} shows the time overhead for clock transition steps as a function of the active frequency.
The execution order matters as save and update are first run at the initial frequency while for re-init the target frequency is already in effect.

In detail, we find that 
the actual frequency change incurs significantly less overhead than post-processing (\ie re-init).
This suggests that most impactful performance optimizations should focus on code for peripheral re-initialization.
Driver init-code is also not likely to be optimized for execution time, since with static clock configuration it is only executed at boot.
In many cases re-init steps can be avoided by using independent clock domains for peripherals or by ensuring that the same frequency is maintained for peripherals.
It is worth noting that the save step is only needed for the current implementation to copy the previous state of configuration registers into the memory.

 	\section{Case Study: Energy-Efficient Networking}
\label{sec-networking}

\newcommand\measurevar[2]{\csname DATA#1#2\endcsname}

\newcommand\loadfiletoarray[2]{
\readarraysepchar{=}
\readdef{fig/#1}{\data}
\readarray*\data\MyDat[-,2]
\newcounter{datacount#2}
\setcounter{datacount#2}{0}\whiledo{\value{datacount#2} < \MyDatROWS}{\stepcounter{datacount#2}\expandafter\xdef\csname DATA#2\MyDat[\arabic{datacount#2},1]\endcsname{\MyDat[\arabic{datacount#2},2]}}
}

\loadfiletoarray{UDP_send_64_bytes.pdf.measurestats}{send64}
\loadfiletoarray{UDP_recv_64_bytes.pdf.measurestats}{recv64}
\loadfiletoarray{UDP_send_pktsz_vs_rel_energy_vs_topo.pdf.measurestats}{sendtopocmp}
\loadfiletoarray{UDP_recv_pktsz_vs_rel_energy_vs_topo.pdf.measurestats}{recvtopocmp}

\newcommand\roundmeasurevar[2]{\pgfmathparse{\measurevar{#1}{#2}}\pgfmathprintnumber[precision=0]{\pgfmathresult}}
\def\siStatRound#1#2#3{\roundmeasurevar{#1}{#2}~\SI{}{#3}}

Communication is central for \ac{IoT} nodes.
Whenever bulk data (\eg firmware updates) or live data (\eg health parameters) are transmitted, a significant share of the system energy is spent for communication, even if advanced network architectures such as edge processing are deployed.
Systems with static clock configuration introduce a major mismatch between available \ac{CPU} processing speed and (relatively low) throughput needed during transmissions.
We now analyze performance benefits for senders and receivers while \sname optimizes clock~speeds for  networking tasks.

\paragraph{Basic setup}
We implement a plain UDP sender-receiver scenario between two nodes. The sender transmits 64 packets of preconfigured payload to the receiver via a single link.
We disable link-layer retransmissions (\ie ACK requests) and back-off
mechanisms (\ie CSMA/CA) at the sender side to prevent blocking by the lower layer and isolate non-deterministic effects of the environment.
This effectively maximizes the throughput towards the bandwidth-limited radio (\ie minimizes dynamic optimization potential) and therefore represents a pessimistic scenario.
When measuring the impact at the receiver, we enable both mechanisms to maximize the incoming channel throughput at the receiver side.
We conduct 64~runs of each of the following experiment settings and present (converged) averages.

We evaluate \ac{CPU} power consumption and performance impact while applying \ac{DVFS} via \sname.
As in our prior evaluations (see \autoref{sec-eval}), we conduct our experiments on the \textit{nucleo-l476rg} evaluation board, extended by an AT86RF233 IEEE802.15.4 radio module that is connected via \ac{SPI}.
The RIOT firmware includes the default network stack \textit{gnrc} on top of the \textit{at86rf2xx} radio driver in its default configuration with an \ac{SPI} target frequency of \SI{5}{\mega\hertz}.
Since lower \ac{CPU} frequencies allow for the reduction of flash wait states and core voltage but both settings sometimes mutually exclude each other, we also investigate the impact of two policies that favor either fast flash access (FF) or low voltage (LV).

\begin{figure*}
	\centering
	\subfloat[Sending: Reducing core clock frequency significantly improves energy consumption at minor increase of transmission time.]{\includegraphics{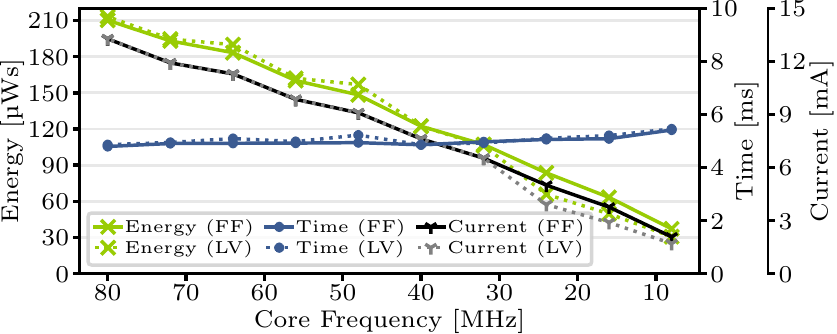}\label{fig_UDP_send}}
	\quad
	\subfloat[Receiving: Reducing core clock frequency saves less energy and exposes higher performance penalty at low core frequencies.]{\includegraphics{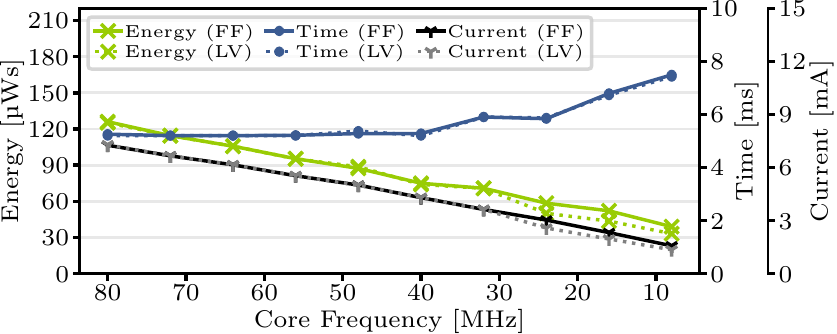}\label{fig_UDP_recv}
	}
	\caption{Using different clock frequencies for sending and receiving 64 bytes payload based on UDP/6LoWPAN.}
	\label{fig_UDP_networking}
	\postfigspace
\end{figure*}

\paragraph{Impact of core clock frequency}
In this experiment, we vary the core clock frequency. \autoref{fig_UDP_send} depicts a strong correlation between current and frequency but a weak correlation between frequency and transmission time (\ie throughput). This indicates significant energy savings at very low (temporal) performance penalty.
When reducing the frequency from \siStatRound{send64}{fmax}{\mega\hertz} to \siStatRound{send64}{fref}{\mega\hertz}, the consumed energy reduces by up to \siStatRound{send64}{energy_decrease_perc_fmax_fref}{\percent} -- whereas the time only increases by  \siStatRound{send64}{time_increase_perc_fmax_fref}{\percent}.
At 1/10th of the frequency (\siStatRound{send64}{fmin}{\mega\hertz}), energy is reduced by $\approx$\,\siStatRound{send64}{energy_decrease_perc_fmax_fmin}{\percent} at a moderate timing penalty of $\approx$\,\siStatRound{send64}{time_increase_perc_fmax_fmin}{\percent}.
This result is line with our expectation because the radio imposes a bottleneck where the packet processing becomes negligible in relation to the time required for transmission.
The LV and FF policies have no significant impact on transmission times, but the low voltage variant is able to further reduce energy consumption.
The fact that network operations do not incur much flash access explains why lower voltage excels faster flash.

In contrast to the sender side, the receiver (\autoref{fig_UDP_recv}) exhibits an overall lower consumption as it effectively requires fewer interactions with the radio.
The \ac{CPU} gets notified asynchronously by the radio once data is available, and can then read and process the packet.
Setting the CPU frequency to half reduces the consumed energy by $\approx$\,\siStatRound{recv64}{energy_decrease_perc_fmax_fref}{\percent} -- again at a very small temporal performance penalty of \siStatRound{recv64}{time_increase_perc_fmax_fref}{\percent}.
The relative energy savings for slowing down to 1/10th are slightly smaller ($\approx$\,\siStatRound{recv64}{energy_decrease_perc_fmax_fmin}{\percent}), which is partially related to the significantly bigger performance impact of $\approx$\,\siStatRound{recv64}{time_increase_perc_fmax_fmin}{\percent} increase in time.
Packet processing only starts after a complete reception, which---if decelerated---stretches the time to become ready for the next reception by freeing the frame buffer. This effectively causes more retransmissions.
The sender runs at its default high core frequency and hence does not reduce stress towards the receiver.
Coordinating clock management across nodes could potentially improve this further.
In this paper we focus on the case of an unaltered environment, as in practice node performance is not assumed to be uniform and protocols must be able to cope with that.

These results clearly show that with \sname both communication directions can leverage unused optimization potentials without sacrificing performance.

\begin{figure*}
	\centering
	\subfloat[Sending]{\includegraphics{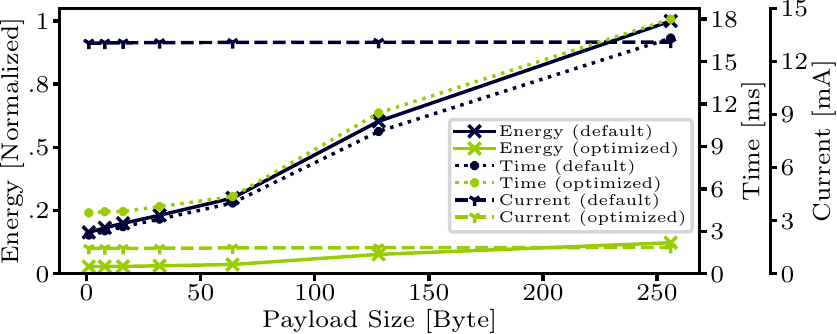}\label{fig_UDP_send_pktsz_vs_energy}}
	\quad
	\subfloat[Receiving]{\includegraphics{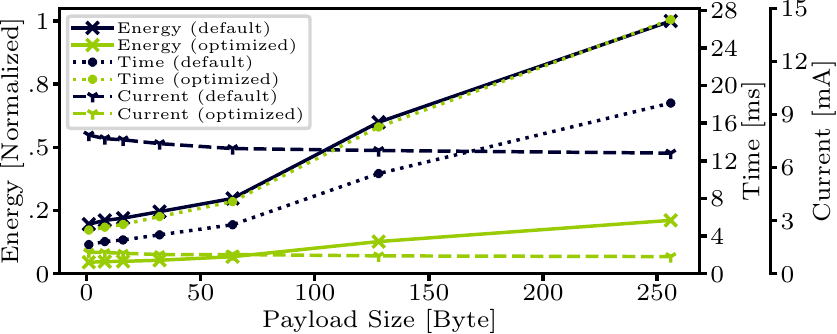}\label{fig_UDP_recv_pktsz_vs_energy}
	}
	\caption{Impact of different payload sizes on energy consumption and timing, comparing the default (high frequency) configuration and the most energy efficient configuration. Energy is normalized to highest consumption.}
	\label{fig_UDP_pktsz_vs_energy}
	\postfigspace
\end{figure*}

\paragraph{Impact of payload size}
The potential for energy savings also depends on the packet size.
 \autoref{fig_UDP_send_pktsz_vs_energy} shows the relation between energy consumption, transmission time, and average current for different payload sizes.
All metrics of the default case are compared to the energy optimized configuration.
Energy values are normalized to the highest energy value (\ie the unoptimized transmission of the biggest payload of \SI{256}{\byte}).
Notably across all packet sizes, the transmission times increase far less than the energy consumptions reduce.
One reason for this is that even the smallest packet still induces a relevant amount of communication towards the radio module, which is also throughput limited because of the \ac{SPI} bus.
Additionally, even a single byte of payload comes with several bytes for involved communication protocols.
The average current is flat for the default and the optimized variant, uncorrelated with the packet size.
In the optimized case the relative time penalty grows towards the smallest payload (1 Byte) compared to higher payloads as the processing time (limited by the slower CPU speed) becomes a more dominant factor.

Three significant differences become apparent at the receiver-side (see \autoref{fig_UDP_recv_pktsz_vs_energy}):
\one the smaller average current consumption across the full set of measurements;
\two the average current now noticeably increases for very small packets, which declines slower for bigger payloads;  
\three the temporal performance is overall more affected by the energy optimization.
This is directly reflected by the level of energy savings in the optimized case, clearly visible when comparing the values of bigger payloads to the sender case.

\begin{figure*}
	\centering
	\subfloat[Sending]{
    	\includegraphics{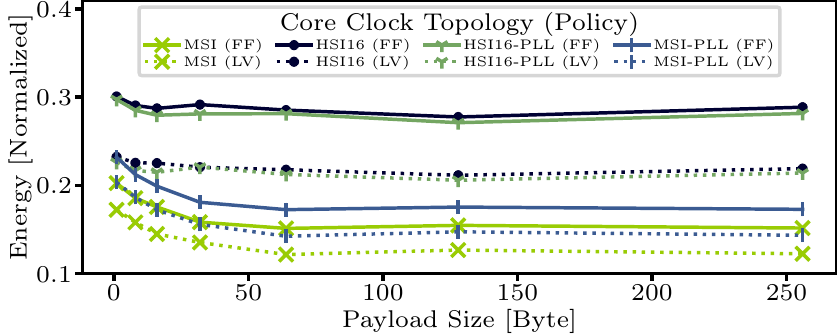}
		\label{fig_UDP_send_pktsz_vs_rel_energy_vs_topo}
	}
	\quad
	\subfloat[Receiving]{
		\includegraphics{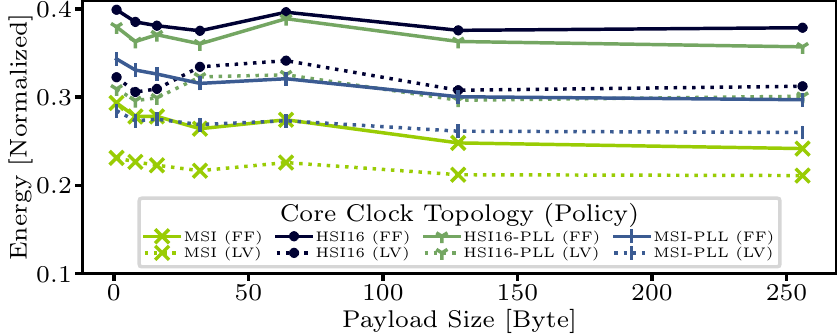}
		\label{fig_UDP_recv_pktsz_vs_rel_energy_vs_topo}
	}
	\caption{Energy impact of payload sizes, comparing the default  (high frequency) and different topology configurations.}
	\label{fig_UDP_pktsz_vs_rel_energy_vs_topo}
	\postfigspace
\end{figure*}

\paragraph{Impact of clock topology}
Last, we investigate the effect of adapting clock topology, \ie the different clock paths used to derive a specific frequency.
\autoref{fig_UDP_send_pktsz_vs_rel_energy_vs_topo} compares the energy for transmission with different core clock topologies using either LV or FF policy.
Here, all energy values are normalized to the energy consumption of transmitting the same payload while operating at the default core clock topology (in this case \texttt{MSI-PLL} at static frequency of \SI{80}{\mega\hertz}).
In all cases, the energy efficiency improvements are slightly lower or very small payloads.
It is possible to achieve significant energy savings of up to \SI{70}{\percent} with FF and more than \SI{77}{\percent} with LV policy for any topology setting.
Yet, being able to switch the topology improves energy efficiency by at least another \siStatRound{sendtopocmp}{worst_to_best_topo_energy_diff_min}{\percent} and up to \siStatRound{sendtopocmp}{worst_to_best_topo_energy_diff_max}{\percent} when comparing to only adapting frequency via the top-most topology.

\autoref{fig_UDP_recv_pktsz_vs_rel_energy_vs_topo} shows the equivalent results for the receiver side.
The relative energy savings are slightly lower due to the generally lower current, as we already observed in \autoref{fig_UDP_recv}.
This partially derives from a reduced bus communication with the radio module because of asynchronous triggering of processing steps.
In contrast, the sender requires several steps that poll the radio for register state changes via the \ac{SPI} bus.
Different from the sender, there are also cases where different topology configurations outperform others.
This can be seen for payloads below \SI{50}{\byte} (\texttt{HSI16} vs \texttt{MSI-PLL}) and hints at certain dynamic application conditions that may change which topology is most preferable.

Overall, these results bolster our assumption that versatile dynamic clock reconfiguration is valuable for leveraging the full energy optimization potential of low-power hardware.

 	\section{Related Work}
\label{sec-rel-work}

Even after more than a decade of research on advanced dynamic energy management for constrained devices~\cite{jtotd-aemws-07}, the limit on power consumption keeps getting pushed lower~\cite{kszsk-ntpsc-20}.
More devices leave batteries behind and adopt intermittent power sources \cite{jp-resfr-20, msl-csip-20}.
Adaptability becomes a key feature for optimizing energy consumption under dynamic conditions \cite{brsyh-rfdfh-21} and development paradigms shift towards systems with more sophisticated abstractions \cite{lcggp-mcse-17}.
Following this perspective, we briefly highlight related work that guides future directions which are not yet applicable to our domain.
\ac{IoT} related solutions are then discussed in more detail for the related areas of \ac{DVFS} and energy efficient networking.

Simonovi\'{c} \etal \cite{szs-amctc-16} propose to encode clock trees using a template based formal language, which  eventually could be provided by manufacturers.
Such common representation of clock trees would mitigate the error-prone manual translation from data sheets.
In a case study, they model a rather complex multiprocessor system-on-a-chip but do not share details on the application or quantitative performance results.

The \ac{CCF}~\cite{t-tccf-20} of Linux handles clock configuration via its device tree~\cite{elw-dtr-20}. \ac{CCF} uses dynamic memory allocation, recursive operations, and large data structures, all of which are avoided on constrained \acp{MCU}.
Furthermore, it can not explore configurations.

Liu \etal \cite{lqw-eadvf-08} propose EA-\acs{DVFS}, an energy aware \ac{DVFS} approach designed for energy harvesting systems that run tasks at an appropriate clock speed depending on energy availability.
The authors evaluate their approach on a rather powerful embedded system with a processor running up to \SI{1}{\giga\hertz}.
They focus on real-time systems where the energy savings reduced deadline misses by more than \SI{50}{\percent}.

\ac{DVFS} continues to trickle down from data centers~\cite{bpopf-ieema-17}, personal computers and smartphones to more constrained target devices like wireless sensors~\cite{kbw-iruws-16, acilp-idvfs-17,zmxy-eicgc-17}.
\newcommand{\ddvfs}{D$^{2}$VFS\xspace}
Chiang \etal \cite{cagll-pcdmc-21} propose a dynamic clock management system that aims for power reduction of tasks limited by IO operations and switches between distinct clocks for that purpose.
Different to our work, they use an implicit mechanism that manages the active clock. We employ a proactive online assessment to provide fine-grained, energy-aware processing adaptations for tasks.
Further, we consider an abstract clock model, advanced topology control via complex transitions, and voltage scaling in addition to frequency scaling.

Ahmed \etal \cite{aasma-icdvf-20} propose \ddvfs, a discrete \ac{DVFS} variant for the intermittent device class.
It scales down frequency and voltage as the supply capacitor empties during operation to increase clock cycles available for processing.
In contrast to \sname it limits possible frequencies to a small subset, instead of providing full control over the clock tree.
Our proactive assessment method steers system performance towards given optimization goals. \ddvfs is reactive and does not aim for a generic clock configuration architecture.

Kulau \etal \cite{kbw-iruws-16} propose IdealVolting, which leverages safety margins of the manufacturer voltage level specification by undervolting far below the specification limits, enabling energy savings of more than \SI{40}{\percent}.
In this work, we do not undervolt below recommended specification limits and employ hardware with integrated voltage scaling capability.

Antonio \etal \cite{acilp-idvfs-17} introduce a programmable power management on \ac{WSN}-class devices that controls \ac{DVFS} and power gating on the chip level.
Their work is orthogonal to ours, as they focus on automatic power gating and \ac{DVFS} implemented in hardware by a custom design on the register-transfer level.
\sname instead improves software control for devices that already provide those features.
Since such devices are becoming more widely available~\cite{kackz-sadpn-18}, we investigate how these features can be uniformly exposed to upper software layers. We want to bridge the gap between specific hardware capabilities and energy-aware software which is platform-agnostic.

Improved energy efficiency of low power radio communication was approached on higher layers with topology control algorithms \cite{mgzn-eeltc-07}, and the physical layer via ultra low power wake up radios \cite{pmktb-ulpwr-17}, or polymorphic radios that dynamically adapt between active and backscatter transmissions~\cite{rgkg-prndp-18}. As our work exploits energy savings related to the common fundamental mismatch between \ac{CPU} and radio throughput, we expect systems with such ultra low-power radios to also significantly benefit from our solution.

 	\section{Conclusion and Outlook}
\label{sec-conc}
In this paper, we proposed \sname, an approach for generic online clock reconfiguration that suits constrained \ac{IoT} devices. 
We reduced the complexity of managing clock dependencies by modeling clock trees as simple, reusable base components with a memory efficient way for encoding configuration parameters and constraints.
Based on this lean model, we could enable dynamic exploration and reconfiguration of clock trees at runtime.
We demonstrated the validity of our concept by implementing \sname on two independent hardware platforms and evaluated its significant impact on energy savings when used for cross-platform \ac{DVFS}.

With \sname we have a tool at hand that allows for fine-grained control of core system parameters. This may help to master emerging challenges. 
Future directions of this research are threefold.
First, additional control feedback mechanisms should be investigated across manifold application scenarios, including complex IoT networks and services.
Second, our algorithms shall be studied in a supervised long-term deployment.
Third,  our generalized, hardware-agnostic access to the clock configuration shall give rise to new use cases and applications of timekeeping and signal generation.

\noindent
\textbf{Artifacts:} 
All artifacts are available openly on GitHub \url{https://github.com/inetrg/RIOT/tree/ScaleClock}

\noindent
\textbf{Acknowledgements:} Funding was provided by the Hamburg \emph{sharing.city.college} of \emph{ahoi.digital} and the BMBF project \emph{PIVOT}.
 	\balance
	\bibliographystyle{IEEEtran}
	\bibliography{own,iot,local,comparch,internet}
	\label{lastpage}	
\end{document}